# Enhancing User Experience in Virtual Reality with Radial Basis Function Interpolation Based Stereoscopic Camera Control

Emre Avan, Ufuk Celikcan, Tolga K. Capin, and Hasmet Gurcay

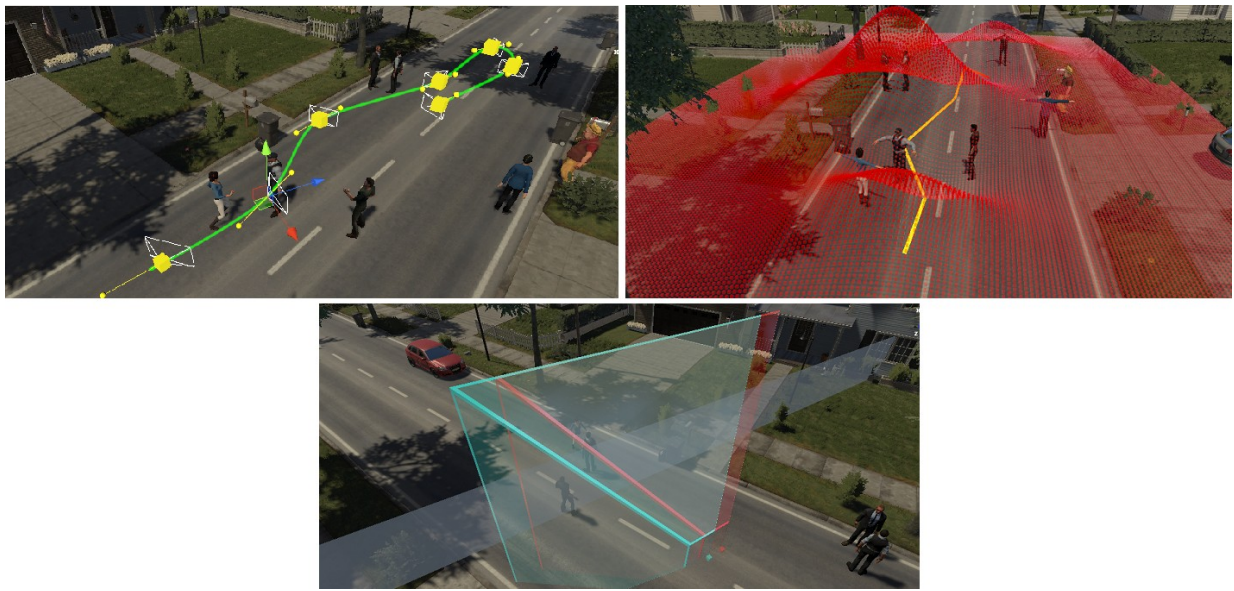

Fig. 1: An overview of our approach. Clockwise from the top left: (1) The user creates a unique design of stereoscopic cinematography, which is composed of a series of waypoint locations with designated camera angles and stereoscopic camera parameters. (2) Then, the method calculates a smooth camera path and interpolates the stereoscopic camera parameters for the rest of the scene using radial basis functions. (3) Finally, during the VR experience, automated stereoscopic camera control is realized on the path by projection matrix manipulations using the fitted surfaces of parameters.

**Abstract**— Providing a depth-rich Virtual Reality (VR) experience to users without causing discomfort remains to be a challenge with today's commercially available head-mounted displays (HMDs), which enforce strict measures on stereoscopic camera parameters for the sake of keeping visual discomfort to a minimum. However, these measures often lead to an unimpressive VR experience with shallow depth feeling. We propose the first method ready to be used with existing consumer HMDs for automated stereoscopic camera control in virtual environments (VEs). Using radial basis function interpolation and projection matrix manipulations, our method makes it possible to significantly enhance user experience in terms of overall perceived depth while maintaining visual discomfort on a par with the default arrangement. In our implementation, we also introduce the first immersive interface for authoring a unique 3D stereoscopic cinematography for any VE to be experienced with consumer HMDs. We conducted a user study that demonstrates the benefits of our approach in terms of superior picture quality and perceived depth. We also investigated the effects of using depth of field (DoF) in combination with our approach and observed that the addition of our DoF implementation was seen as a degraded experience, if not similar.

✦

## 1 INTRODUCTION AND MOTIVATION

In recent years, Virtual Reality (VR) has become more commonplace with recent advances in hardware technology that have led to the production of consumer appropriate head-mounted displays (HMDs), such as HTC Vive and Oculus Rift. A wide field-of-view (FOV) HMD that immerses a user in computer-generated virtual worlds is a key enabling technology to VR applications.

The immersive nature of HMDs creates a strong presence illusion, where users perceive virtual environments (VEs) as real and not mediated through technology. Therefore, it finds a myriad of applications in gaming, entertainment, simulation and training, defense, education, and other fields. Support from the makers of the commercially available HMDs with extensive software development kits has resulted in an unseen and rapidly expanding ecosystem of such applications specifically designed for VR. As VR has been increasingly accessible and popular, comfortable, high-quality stereo 3D has become an important and timely requirement for real-time VR applications. There are still issues that remain to be resolved in order to provide a thoroughly realistic and comfortable experience to VR users. Immersive VEs can be visually constraining. The foremost contributing factor to visual discomfort is the accommodation-convergence conflict (ACC), which arises due to

- *Emre Avan is with Hacettepe University, Department of Computer Graphics. E-mail: emreavan@gmail.com.*
- *Ufuk Celikcan is with Hacettepe University, Department of Computer Engineering. E-mail: ufuk.celikcan@gmail.com.*
- *Tolga K. Capin is with TED University, Department of Computer Engineering. E-mail: tolga.capin@tedu.edu.tr.*
- *Hasmet Gurcay is with Hacettepe University, Department of Mathematics. E-mail: gurcay@hacettepe.edu.tr.*

the dissonance between accommodation, adjustment of the eye lenses to focus at the observed depth, and eye convergence [21, 25, 34, 39]. While these two cues are cross coupled in normal viewing conditions, in stereoscopic displays, the viewer always focuses at the screen level regardless of where eyes actually converge, which leads to ACC. Due to the ensuing discomfort, users commonly report symptoms such as eyestrain, nausea, dizziness and headaches after using HMDs for extended periods. Enhancing user experience and perceived depth together without invoking discomfort has been a major challenge in stereoscopic content production.

To create stereo vision for VR applications, there are two main parameters that need to be set. These are interaxial separation, which is the distance between the two cameras, and their convergence distance in the VE. These stereoscopic camera parameters play a major role in the VR experience as they produce the disparity between left and right images, and therefore impact the amount of perceived depth [16], as well as visual comfort. Nonetheless, no single setting exists that minimizes the fusion effort and leads to optimized depth perception for varying viewing circumstances and depth ranges [24].

In current commercially available HMDs, stereo camera parameters are set to be fixed at the API level, such that, they are kept constant no matter how the scene contents change or the depth composition of the scene varies. This is a convenient solution to avoid issues regarding visual discomfort, however it reduces the depth perceived, and, in turn, level of immersion of the users.

In this work, we aim to enhance user experience in VR with consumer HMDs in terms of overall perceived depth without sacrificing picture quality or visual comfort. Addressing the challenges of changing depth composition dynamically while maintaining visual discomfort to a minimum, we propose a new method for automated stereoscopic camera control in VEs. To the best of our knowledge, this is the first such method ready to be used with existing consumer HMDs without additional hardware requirements such as embedded eye trackers or focus-adjustable lenses.

Our proposed method demands a stereoscopic cinematography, that is, a particular arrangement of stereoscopic camera settings consisting of a series of waypoint locations with designated camera angles and stereoscopic camera parameters. The stereoscopic parameters associated with these locations constitute a set of scattered data. Using radial basis function interpolation on this small set of data, our method produces a smooth surface fit of parameters for the rest of a given VE and provides automated stereoscopic camera control by continuous projection matrix manipulations using the fitted parameters.

We also introduce a VR interface for creating the required stereoscopic cinematography. Our survey of the relevant literature indicates that this is the first immersive interface that can readily be used with all commercially available HMDs for authoring a unique 3D stereoscopic cinematography. With the proposed interface, users can author unique depth narratives for any given VE directly from the first-person perspective exactly as the VE will be experienced using the same HMD that it will be experienced with. While it can be used by VR content creators to design signature depth narratives, the easy-to-use interface lets even the most novice users to quickly create their own unique VR experiences.

As the highly subjective nature of stereoscopic camera control necessitates, we evaluated our method in comparison to the default HMD settings in different configurations. The results illustrate that our method is able to significantly enhance user experience in terms of overall perceived depth while also boosting picture quality and maintaining visual discomfort on a par with the default arrangement. We also analyzed whether using Depth-of-Field blur in addition to our method improves user experience further and found that, as in some similar earlier studies, the addition of our Depth-of-Field blur implementation was seen as a degraded experience notably in terms of visual comfort and similar in terms of depth perception.

The remainder of this paper is organized as follows. In Sections 2 and 3, we briefly give some background information on the components of our work and present an overview of previous works on the subject matter. Our proposed method is elaborated in Section 4. Then, in

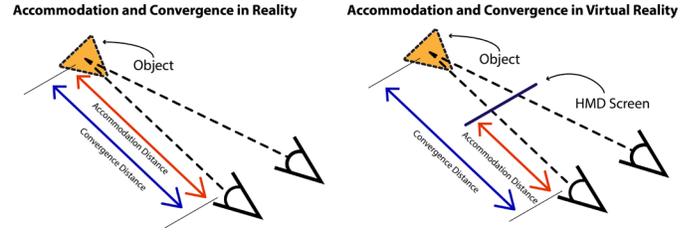

Fig. 2: Cross coupling between accommodation and convergence is broken in stereoscopic display systems such as HMDs.

Section 5, the details of our user evaluation study are given and its results are illustrated and discussed. Finally, Section 6 concludes the paper.

## 2 BACKGROUND

### 2.1 Stereo Vision and Stereoscopic Rendering

*Interaxial separation* and *zero-parallax distance* are the two major parameters in stereoscopic rendering. For the binocular vision, there are two virtual cameras rendering the scene with a slight distance between each other. This distance is called interaxial separation and lets the human visual system to create depth perception. Secondly, oculomotor muscles enable eyes to converge into a plane which is called *zero-parallax plane*, also known as convergence plane, and the distance between this plane and the cameras is called zero-parallax distance.

If the object is far behind the zero-parallax plane, it has positive disparity and appears as inside the display. For an object that is at a closer distance than the zero-parallax plane, it appears as in front of the display and this situation is called negative disparity. As the zero-parallax distance gets closer to the virtual cameras, image for the left eye shifted to left and right for the right eye for the HMDs. For the objects that have negative disparity, fatigue may occur more easily than the ones with positive disparity since oculomotor muscle needs to contract in order to rotate eyes inwards to focus with negative disparity. Therefore, developers and designers carefully control the scene composition and cinematography in order to keep users' eyes rested as much as possible [13].

When looking at the screen, viewer's eyes converge or diverge according to the depth of the object in the scene while they are focused on the display. In real-world, accommodation and convergence systems are cross-coupled, which means that eyes both converge and accommodate at the same position as seen in the Fig 2. Crystalline lens of the eye make accommodation possible by refracting the light inwards or outwards. In stereoscopic display systems, on the other hand, coupling between accommodation and convergence is broken causing ACC.

In order to solve this issue which can be seen in Figure 2, focus object needs to be in stereoscopic comfort zone. This zone is defined by Percival [31] to be 1/3 diopter distance from each side (negative disparity and positive disparity) to the accommodation distance. To provide the salient VE contents within stereoscopic comfort zone, the perceived depth should be limited [41].

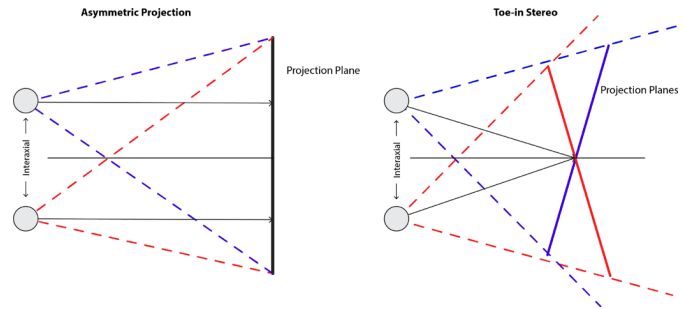

Fig. 3: Difference between Asymmetric Frustum (on the left) and Toe-in (on the right) camera setups.

Typically, in plano-stereoscopic displays like 3D TV or cinema screen, where disparity often falls within the comfort zone as viewers observe the images at a large distance, image distortion is not a severe issue.But this is not the case for HMDs since the display is placed at a very close proximity to the eyes. Unnatural stereo images can cause conflicting cross-coupled interactions between the crystalline lens and oculomotor nerve systems which result in discomfort [27] and can induce VR sickness. Moreover, selected depth should be changed smoothly. Sudden changes in zero-parallax distance between near and far regions can be disorienting [40].

There are two common methods rendering the scene with a given convergence distance. One is stereo camera rotation, which is a relatively easier to develop solution than the former method for disparity range adaptation. This method is also known as toe-in stereo and it is widely used for stereoscopic contents that is viewed at stereoscopic displays. In this method, both virtual cameras are rotated according to the zero-parallax plane distance. However, rotating cameras creates distorted images which is called Keystone distortion [4]. Vertical asymmetry occurs at the corners of the screen, since virtual cameras look at different axis which is shown at in Figure 3. HMDs such as HTC Vive and Oculus Rift also use the asymmetric projection matrix where both cameras look outwards in order to shift the scene contents inwards.

The other method is called asymmetric projection [4], also known as off-axis projection, which is used by current consumer HMDs. With this method, the projection matrix is calculated to make cameras have an asymmetric frustum, such that, one side of the view frustum has more area to be rendered, making one side of the view frustum more condensed and the other side expanded. Since the vertical asymmetry is avoided, keystone distortion does not occur in asymmetric projection.

## 2.2 Radial Basis Functions

Radial Basis Functions (RBFs) are identified as one of the most accurate and stable methods of solving problems involving scattered data interpolation [11, 32, 33]. It is possible to obtain an implicit surface representation approximating the scattered data using RBFs.

RBF is a radially symmetric function centered around a point $x_c$. For a collection of scattered data $\mathbb{X} = \{x_1, x_2, \ldots, x_n\}$ of $n$ distinct data points and a matching collection of values $y_1, y_2, \ldots, y_n$ sampled from an unknown function $f$, s.t. $y_i = f(x_i)$, we can select an RBF $\Phi$ and a group of centers $\{x_{c_1}, x_{c_2}, \ldots, x_{c_m}\}$, $m \in \mathbb{N}$, to form a basis $\{\Phi(\|\cdot - x_{c_1}\|), \Phi(\|\cdot - x_{c_2}\|), \ldots, \Phi(\|\cdot - x_{c_m}\|)\}$. Then, this basis can be utilized to construct an approximation $\widetilde{f}$ of $f$, such that there will be one basis function $\forall i\{1, 2, \ldots, n\}$ where $x_i$ is the center $x_c$ of that function. Thus, the approximate function $\widetilde{f}$ can be assembled with those $n$ RBFs as

$$\widetilde{f}(x) = \sum_{j=1}^{n} w_j \Phi(\|x - x_j\|)$$

with constant $w_j$.

RBFs have been utilized in various other areas of computer graphics. Carr et al. used RBFs to reconstruct smooth, manifold surfaces from point-cloud data [7]. In the work by Noh et al., RBFs were utilized for creating deformations of polygonal models to produce animations by localized real-time deformations [26]. Weissmann and Salomon used RBFs in gesture recognition for VR applications [35]. Yet, to the best of our knowledge, there has not been a work making use of RBFs for stereoscopic rendering.

## 3 Previous Works

The main concern of content creators while designing a virtual environment for stereo displays is to keep the salient contents within the comfort zone where user can experience the scene without discomfort while maintaining the depth perception. In VEs where the camera is in constant motion, a control system is needed to keep the perceived depth range within the visual comfort zone. Lang et al. [22] proposed a stereoscopic camera controller for post-production disparity range adjustment while Oskam et al. [29] introduced a controller for real-time disparity range adaptation by optimizing stereoscopic camera rendering parameters. Celikcan et al. [8] suggested another real-time optimization based controller that addresses the interaction of binocular depth perception and object saliency by factoring viewer's individual stereoscopic disparity range into the optimization. However, these methods were developed for plano-stereoscopic displays and have not been adapted to work with HMDs.

In the early works, focus adjustable lenses for dynamic convergence or monovision are proposed to solve ACC. Focus adjustable lenses allow users to accommodate accurately and minimize decoupling between accommodation and convergence. However, focus adjustable lenses are still in development and it is not possible to simply use them as add-ons with current consumer HMDs. Monovision, on the other hand, uses two different focal point lenses to let one eye to focus near while the other focuses far distances. According to the studies [19, 30], while monovision reduces fatigue, eye irritation and headache, viewer comfort is not improved.

There have been various works suggesting the use of Depth-of-Field (DoF) blur in stereo vision systems as a remedy for visual discomfort. With DoF blur effect, regions that are far away from the focus plane are blurred to simulate how human visual system perceives objects in full sharpness only within some range round the focal distance.

O'Hare et al. [28] suggested that DoF blur can be used to supplement perceived depth. It is also noted that the effect would only be useful when the blur gradients do not cause discomfort themselves. If blurred areas contain objects that are salient for the viewer, it will degrade viewer's experience and cause even more discomfort. Koulieris et al. [18] also suggest the use of DoF rendering towards achieving goals such as realistic sense of depth and adjusting the perceived scale.

According to Carnegie and Rhee [6], unlike plano-stereoscopic displays, people rather use head motions instead of eye motions for spatial movement and tend to look at the center of the display while experiencing VEs with HMDs. Therefore, it is suggested that using DoF blur to keep the center of the screen in focus will reduce discomfort for many users. In their study, participants felt less discomfort while experiencing the scene with DoF blur.

On the other hand, Langbehn et al. [23] reported that blur is not crucial for the human visual system to extract depth and motion information according to the findings of their study. They state that the HMDs' narrow FOV and low resolutions are the chief reasons behind this. It is also found that there is a trade-off between visual comfort and sense of presence when blurring is employed; and, therefore, an optimization scheme is advised [27].Conti et al. [9] studied the impact of DoF blur with varying interaxial camera distance. They examined configurations in the presence or absence of DoF blur and found that, as with some earlier similar attempts [5, 10], the configurations with DoF blur obtained lower subjective evaluations.

At any rate, to be able to use DoF blurring approach effectively, it is imperative to have, or at least estimate, user's instant gaze location which is not possible without using a high-precision eye tracker embedded inside the HMD or accurately predicting the most salient region of the frame in real time.

When available, gaze data can also be used for direct disparity manipulations in order to improve visual comfort by reducing ACC, frame violation, and crosstalk [15], and enhance depth perception [17]. However, with the existing commercial stereo displays and HMDs, it is not yet possible to find where the user's eyes converge. Therefore different methods have been proposed to solve ACC where the convergence plane is estimated to be.

To find the zero-parallax distance, Shertyuk and State [36] suggest that human eyes will mostly maintain convergence on the hand as a contact point when performing direct object access and manipulations. With that assumption, they find the target object's position and rotate the cameras accordingly. Summer [37] uses a similar camera rotation approach positioning the zero-parallax plane where the target object at the center of the screen is. Koulieris et al. [19], on the other hand, designed a machine learning gaze predictor based on game variables and eye tracking data from a video game. After finding the zero-parallax plane, off-center projection matrix shifts that plane into the comfort zone.

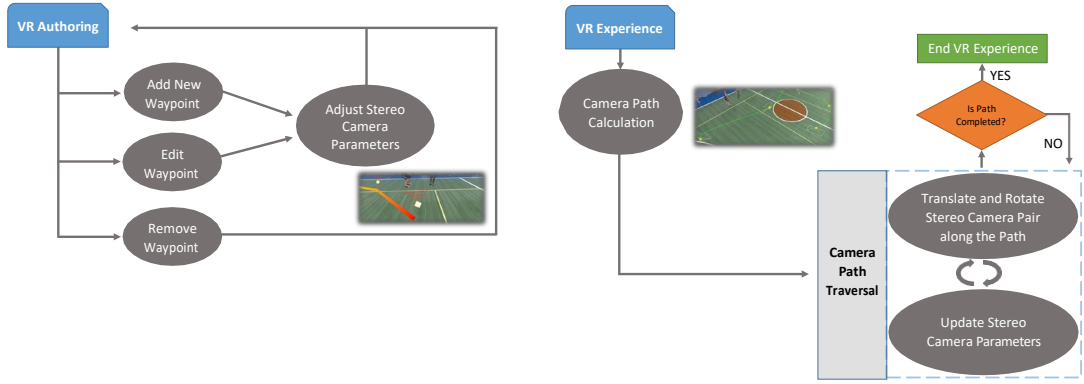

Fig. 4: Flow diagram of our RBF interpolation-based stereoscopic camera control approach.

While some gaze prediction-based methods have been shown to be effective at improving VR experience of users within some restricted settings, each has certain shortcomings such as limited usability, low accuracy or low frame rate. While equipping HMDs with eye trackers can solve most of these issues by foveated rendering, they are not available with current consumer HMDs, with the exception of HTC Vive Pro Eye, which remains to be a niche product mainly due to its considerably higher price. To the best of our knowledge, our work presented in this paper is the first that does not require gaze detection or prediction and can readily be used with existing consumer HMDs for automated stereoscopic camera control in VEs towards enhancing the VR experience.

## 4 RBF Interpolation-Based Stereoscopic Camera Control

The overview of our RBF interpolation-based stereoscopic camera control approach is shown in Figure 4. The approach consists of two parts, namely, VR Authoring and VR Experience. While VR Experience is the main part where the automatic camera control takes place, it requires an arrangement of stereoscopic camera settings, a *depth layout*, that is created in VR Authoring. Users can promptly start their VR Experience with one of the preset depth layouts or use VR Authoring to create their own depth layouts.

### 4.1 VR Authoring

In VR Authoring, the user designs the *depth narrative* of the VR experience by creating a depth layout. A depth layout is a unique design of stereoscopic cinematography that consists of a path for the stereo camera pair to follow with designated camera angles and stereoscopic camera parameters. A depth layout is composed of a series of waypoints. By placing a waypoint, the user defines a position and an angle for the stereo camera pair and sets the stereoscopic camera parameters.

During authoring, the user is placed in the VE that they are creating a depth layout for. This way, they can tailor a depth narrative to that VE from the first-person perspective in the same immersive setting using the same display (HMD) viewers will experience it with. This aspect of our approach constitutes a significant improvement over the traditional stereoscopic editing paradigm where users are bound to work with a two-dimensional interface on a two-dimensional display.

The user can roam the VE and place a waypoint anywhere in it freely. Movement in the VE is realized by either materially moving in the physical space or virtually teleporting within the VE. With our editing in fist-person view paradigm, position and angle of the stereo camera pair are controlled by the user via HMD in the same way a cameraman operates their camera in live shooting. Upon locking the position and view angle for a waypoint, the user adjusts the stereoscopic

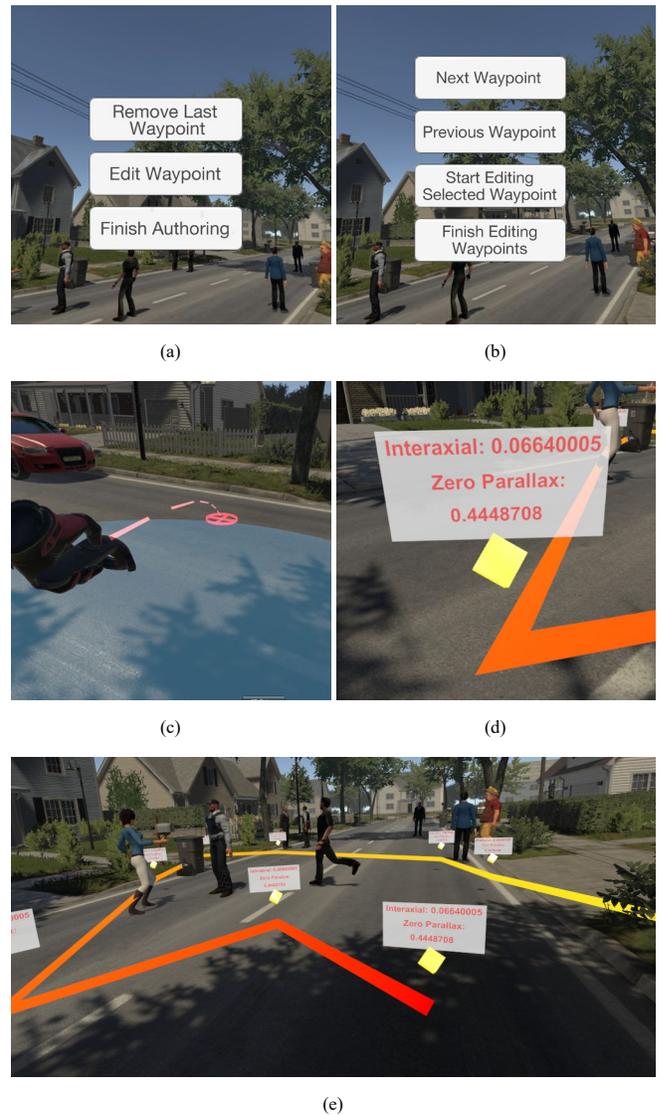

Fig. 5: The graphical user interfaces in VR Authoring: (a) main menu (b) submenu for Edit Waypoint (c) the user is restricted to changing the position of an existing waypoint within a certain radius (d) close-up of a waypoint marker with its saved zero-parallax distance and interaxial separation hovering above (e) preview of a sample camera path.

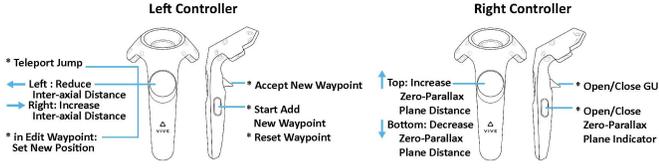

Fig. 6: Users can interact with the system using the hand controllers of HTC Vive. The printed version of this figure is shown to the subjects before the testing starts to have them familiarized with the interface. Hand controller images courtesy of HTC Vive [1].

camera parameters with the corresponding dials on the hand controllers (Fig. 6).

We developed the physical and graphical interfaces for VR Authoring with minimality and ease-of-use as our primary design precepts. Figure 5 shows the graphical user interfaces (GUI) of VR Authoring. At the start of authoring, in accordance with our design precepts, no GUI is displayed to the user so that they can start and continue adding waypoints with minimal interruption until they decide to finish authoring.

At any point during authoring, the user can reveal the GUI main menu (Fig. 5a) by pressing the open/close GUI button on the right hand controller (Fig. 6) and choose to remove the last added waypoint, edit existing waypoints or finish authoring by selecting the corresponding button. If Edit Waypoint is selected, a submenu is shown (Fig. 5b). With the submenu, the user can move between existing waypoints to select one for readjusting its position, view angle or stereo camera parameters. The only difference in editing an existing waypoint than adding a new one is that the user now can change the position of an existing waypoint within a limited radius around the current position (Fig. 5c).

The user can define as many waypoints as they like. After each waypoint is saved, the user can see the actual progress by the preview path with gradient colored lines (Fig. 5e) in-between the placed waypoints markers, which are indicated by cubes and have the saved stereo camera parameters hovering over them (Fig. 5d).

When the user is content with the depth layout, they can exit VR Authoring by pressing Finish Authoring button in the main menu.Then, the user can switch to the VR Experience mode to preview the layout and then return back to authoring for further additions, removals or adjustments of waypoints.

### 4.2 VR Experience

In this mode, our method continuously updates stereoscopic camera parameters, interaxial separation of the stereo camera pair ($D_{ia}$) and zero-parallax distance ($D_{zp}$), while following the camera path. For smooth camera translations, the piecewise linear path designated in the selected depth layout is converted to a Bezier curve using waypoints as the control points. Spherical linear interpolation is used to rotate the camera pair from one waypoint's view angle to the next one's. Head motions coming from the HMD are omitted in order to have the viewers experience the stereoscopic cinematography exactly as authored.

Before the experience starts, for each stereoscopic camera parameter, a separate set of RBF interpolation weights $w_j$, $j = 1,...,m$ are first found by solving the the set of $m$ equations

$$F(x_i) = \sum_{j=1}^{m} w_j \Phi\left(\|x_i - x_j\|\right), \quad i = 1,...,m \tag{1}$$

where $m$ is the number of waypoints, $x_i$ is the position of the waypoint $i$, $\Phi(\cdot)$ is the designated basis function, and $F(x_i)$ is the value of that parameter set for the waypoint $i$ during authoring.

During the experience, for an arbitrary point on the camera path given by the two-dimensional coordinates $x$, the interpolated value $F(x)$

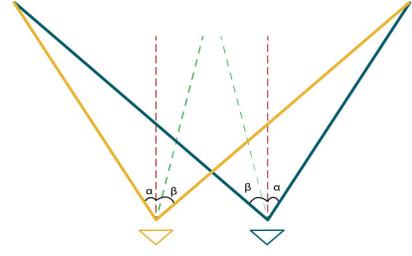

Fig. 7: Asymmetric projection as seen from top-down view. Dashed green line represents the true half angle between left and right planes while dashed red line separates $\alpha$ and $\beta$ angles.

of the parameter is computed by

$$F(x) = \sum_{j=1}^{m} w_j \Phi\left(\|x - x_j\|\right) \tag{2}$$

with respect to the computed weights $w$ and the chosen basis function $\Phi(\cdot)$. We use inverse multiquadric basis function $\Phi(r) = 1/\sqrt{r^2 + r_o^2}$ with $r_o = 2$ in our approach. Inverse multiquadrics are found to provide excellent approximations even when the number of centers $m$ is small [11].

Then, the projection matrices for the two cameras are modified with the updated parameters $D_{ia}$ and $D_{zp}$ based on dynamic projection matrix calculation method for stereo 3D [19]. According to this method, both cameras will have asymmetric projection matrices that together create a single viewing plane at the selected zero-parallax distance. This is suitable for plano-stereoscopic displays since there is only one display for eyes to converge and focus. For the HMDs, on the other hand, there is a dedicated display for each eye rendering using the method as is infeasible. Figure 7 shows the view frusta of the cameras from top-down point of view. Here, when $\alpha$ is larger than $\beta$, it triggers eyes to rotate inwards. As the difference between $\alpha$ and $\beta$ gets larger, the viewer's eyes rotate more and more inwards, which creates the feeling of looking at a nearby object. Leveraging this mechanism, in our approach, the projection matrices for the left and the right eye are swapped, so that, the cameras are directed more outwards as the zero-parallax plane comes closer and less outwards as the plane goes farther to direct viewer attention properly.

Projection matrix can be given as:

$$P = \begin{bmatrix} \frac{2n}{r-l} & 0 & \frac{r+l}{r-l} & 0 \\ 0 & \frac{2n}{t-b} & \frac{t+b}{t-b} & 0 \\ 0 & 0 & \frac{-(f+n)}{f-b} & \frac{-2fn}{f-n} \\ 0 & 0 & -1 & 0 \end{bmatrix} \tag{3}$$

where $l, r, n, f, t$ and $b$ denote left, right, near, far, top and bottom plane positions of the view frustum, respectively. $\frac{r+l}{r-l}$ creates horizontal off-axis asymmetry while $\frac{t+b}{t-b}$ creates vertical off-axis asymmetry [2]. In our approach,we alter the horizontal asymmetry and leave the vertical asymmetry (i.e., $t$ and $b$ values), which is due to the inherent vertical asymmetry of the HMD lenses, unchanged.

Symmetric frustum is used to calculate $l$ and $r$, then half of $D_{ia}$ is added to the expanded side ($l$ to the left virtual camera and $r$ to the right virtual camera) and subtracted from the condensed side at the zero-parallax distance. Hence, in our approach, $l$ and $r$ are calculated as follows. $A$, the half of the frusta's horizontal width at $D_{zp}$, is found as

$$A = D_{zp} \; q \; \tan\left(\frac{\theta}{2}\right) \tag{4}$$

where $\theta$ is the vertical FOV and $q$ is the aspect ratio. At $D_{zp}$, the condensed side's horizontal width $B$ is half of $D_{ia}$ narrower than $A$ and the expanded side's horizontal width $C$ is half of $D_{ia}$ wider than $A$, i.e.,

$$B = A - \frac{D_{ia}}{2} \quad (5)$$

and

$$C = A + \frac{D_{ia}}{2} \quad (6)$$

Since we swap the left and the right view frusta, $l$ and $r$ values are exchanged, i.e., for the left frustum, they are given as

$$l = -C\,\frac{n}{D_{zp}} \qquad r = B\,\frac{n}{D_{zp}} \quad (7)$$

and for the right frustum, they are given as

$$l = -B\,\frac{n}{D_{zp}} \qquad r = C\,\frac{n}{D_{zp}} \quad (8)$$

The new $l$ and $r$ values per view frustum are then plugged in $P$ given in Eq. 3.

Unless additional measures are taken, an important drawback of this method is its sensitivity to $D_{zp}$. Diplopia will occur at large $D_{zp}$ values and small $D_{zp}$ values will cause eye strain and fatigue [13, 20]. In order to remedy this by defining a suitable range for fusion, our system restricts $D_{zp}$ in a dynamic range according to immediate $D_{ia}$ that lets the difference between $\alpha$ and $\beta$ (see Fig. 7) to change within -10° to +1° of the HMD's default.

## 5 EVALUATION

### 5.1 System

The VR framework was realized using Unity graphics development engine. For the RBF-based interpolation, we used the library written in C++ by John Burkardt [3] as a plugin to the framework. During the user evaluations, HTC Vive VR setup was used with a laptop having 3.9 GHz Intel i7 CPU, 16 GB RAM and NVIDIA GTX 1070 GDDR5 8GB GPU. Frame generation rate during the tests varied between 89 and 91 FPS.

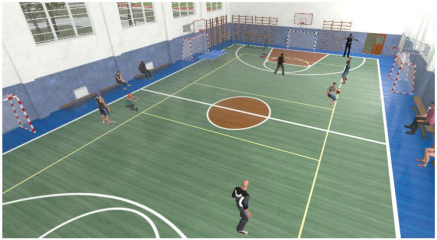

(a)

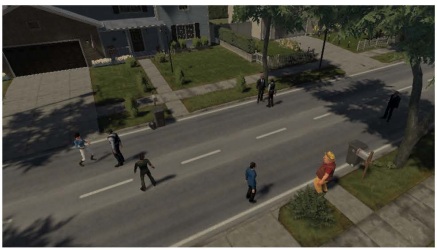

(b)

Fig. 8: Sample frames from (a) the indoor scene (b) the outdoor scene.

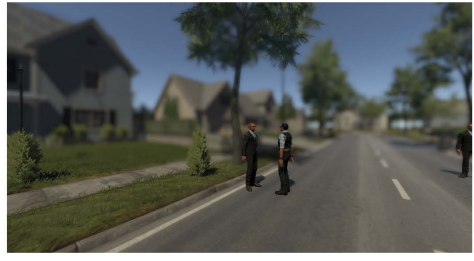

(a)

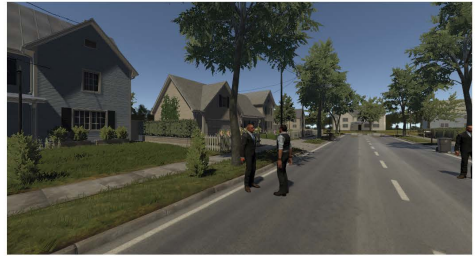

(b)

Fig. 9: Sample frame from the outdoor scene (a) with the added DoF blur starting at zero-parallax distance (b) without DoF blur.

### 5.2 Scenes

We conducted our evaluation using two different scenes. One of the scenes takes place indoors in a school gym with multiple animated athletes and sports fans (Fig. 8a). The other one is an outdoor scene with multiple animated human characters that are situated along a suburban residential street (Fig. 8b).

Maximum render depth is considerably higher in the outdoor scene (see Fig. 12). An often omitted aspect in similar studies, this two-tiered design has the purpose of assessing the impact of available depth range on viewer's perception, e.g., viewer's focus may change more abruptly or frequently between far and near regions when more depth is available and this may cause visual fatigue and eye-strain if the evaluated setting does not offer proper compensation.

### 5.3 DoF Blur Implementation

In our user study, we also test the impact of employing DoF blur on visual discomfort. We used Unity's Post Processing Stack V2 for DoF blur which applies scatter as gather algorithm [14] with multisample anti-aliasing. The focal length and aperture values were set to 75mm and f/5.4, respectively, in our tests. Focal distance of DoF is dynamically matched to zero-parallax distance (Fig.9).

### 5.4 Subjects

We recruited 10 subjects (4 females, 6 males, of ages between 24 and 32 with an average of 28.1) for the tests. All subjects were volunteers and were not compensated for their participation in any form. Before the tests, each volunteer was assessed for proper stereoscopic visual acuity and those who failed did not participate in the evaluation.

The subjects had adequate technical background to use the system. They were not informed about the purpose of the experiment. Only 3 subjects had extensive experience with VR and one of those had done VR application development in the past.

### 5.5 Test Procedure

The subjects were informed about the tasks involved and the key points that they will rate prior to the tests with written and verbal instructions. While authoring was carried out standing, the subjects were seated while they were experiencing the scenes for the sake of comfort.

In VR systems, the initial positioning of the two virtual cameras, which compute the images for the two eyes, is perturbed by the inter-pupillary distance (IPD) setting. This setting is adjusted on the HMD according to the viewer's own IPD, the distance between the centers of their pupils. When the IPD is set incorrectly, perception errors and eye strain may follow [12]. Therefore, at the beginning of the tests, the IPD of a subject is measured using a digital pupilometer and set the IPD value of the HMD correspondingly.

During the tests, our approach was evaluated by the subjects in pairs of sessions for each scene in three different configurations, as given below. Order of the configurations was randomized during the tests.

### 5.5.1 Preset Experience vs. Default Experience

First, the subjects experienced each of the two scenes once in a session with our approach via a preset depth layout (*Preset Experience*) and once in another session with Vive's fixed stereoscopic camera parameters (*Default Experience*). During the Default Experience session, the virtual stereo camera pair follows the same path, with the same camera positions and viewpoints, as the one that was created in the preset depth layout picked for our approach. Order of sessions was randomized at run time and the subjects were not informed about the order.

### 5.5.2 Self-Authored Experience vs. Default Experience

The second configuration is similar to the first one, except this time, the subjects first used VR Authoring to create their own depth layout. They were asked to design two depth layouts in total, one for each VR scene. While the subjects were free to shape their layouts according to their liking, they were told to keep the threshold to a minimum of 5 for the number of waypoints to create within a layout.

Afterwards, they again experienced each scene in two sessions, once with our approach via their self-authored depth layout (*Self-Authored Experience*) and once with the Default Experience setting, in a random order.

### 5.5.3 Preset Experience with DoF blur vs. Preset Experience

In the last configuration, the subjects experienced the two scenes using our approach only, once without DoF blur (Preset Experience, as in the first configuration) and once with DoF blur (*Preset Experience with DoF blur*), via the same preset depth layout (Fig.9). Order of sessions was randomized again without informing the subjects.

### 5.6 Evaluation Criteria

Following each session, the subjects were asked to evaluate the session in terms of image quality, perceived depth and visual comfort in a 5-point Likert scale with the labels "bad", "poor", "fair", "good", and "excellent". These criteria, which are detailed below, are frequently resorted to for perceptual assessment of stereoscopic contents[38].

**Image Quality**: expresses the overall visual quality of the displayed content as perceived by the user. Since our approach dynamically modifies the degree of horizontal asymmetry of the frusta, proper fusion of the resulting left and right images and proper scaling of the scene contents in these images should be established by validating the image quality.

**Perceived Depth**: denotes the apparent depth of the displayed content perceived by the user. With stereoscopic camera control methods aiming at a dynamic depth narrative, as the stereo camera parameters change over time, so does the perceived depth. Accordingly, providing a method that brings upon a feeling of realistic depth is essential for an enhanced VR experience since it contributes a great deal to user immersion.

**Visual Comfort**: is to measure the subjective feeling of visual comfort of the user. Improperly set stereoscopic camera parameters cause visual discomfort in the form of eye strain, which can lead to adverse side effects including visual fatigue, nausea and headaches and yield a dissatisfactory VR experience. Hence, first of all, it is vital to ensure that a proposed stereoscopic camera control method does not invoke visual distress.

Once both sessions and their respective individual assessments are done, the subjects were then asked to evaluate the two sessions vis-a-vis each other. This time, in addition to the previous three, they were also to make a comparison in terms of overall quality.

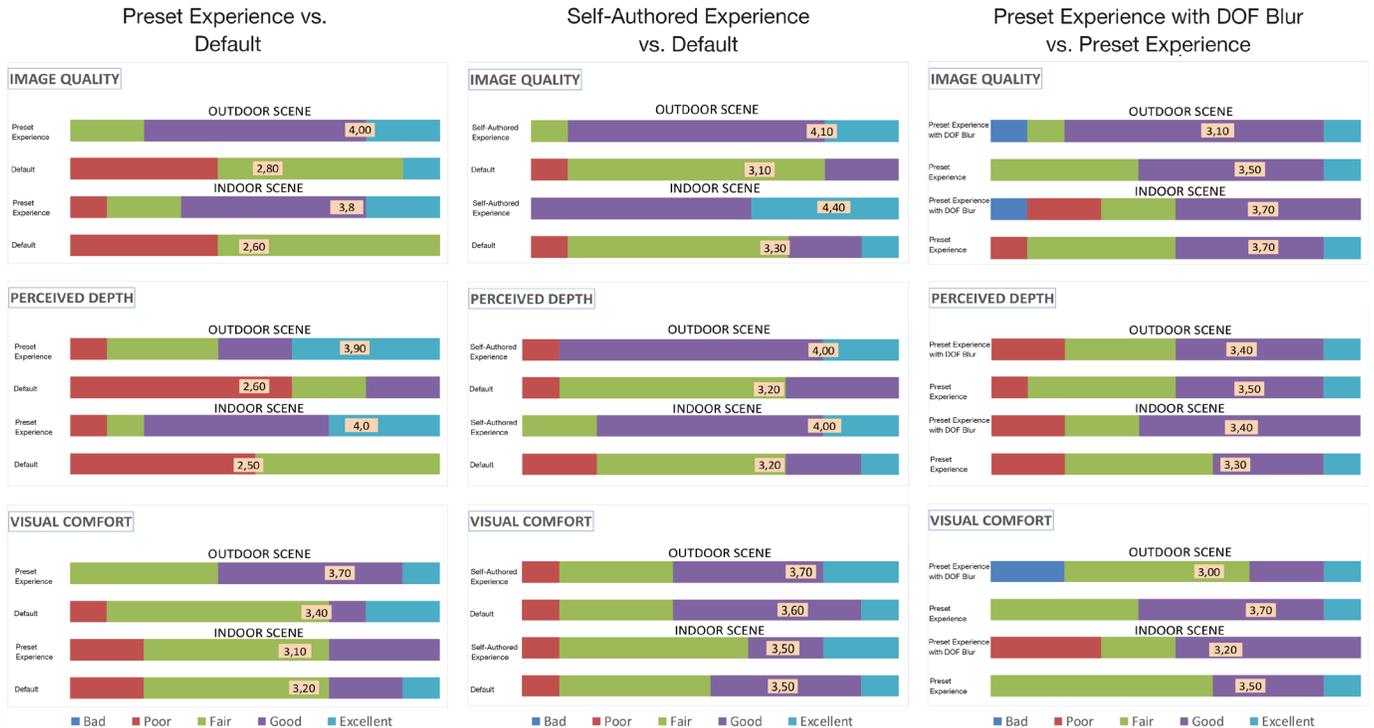

Fig. 10: Individual session ratings collected with the questionnaires and their averages are given for each scene per configuration. The averages are indicated in rectangles.

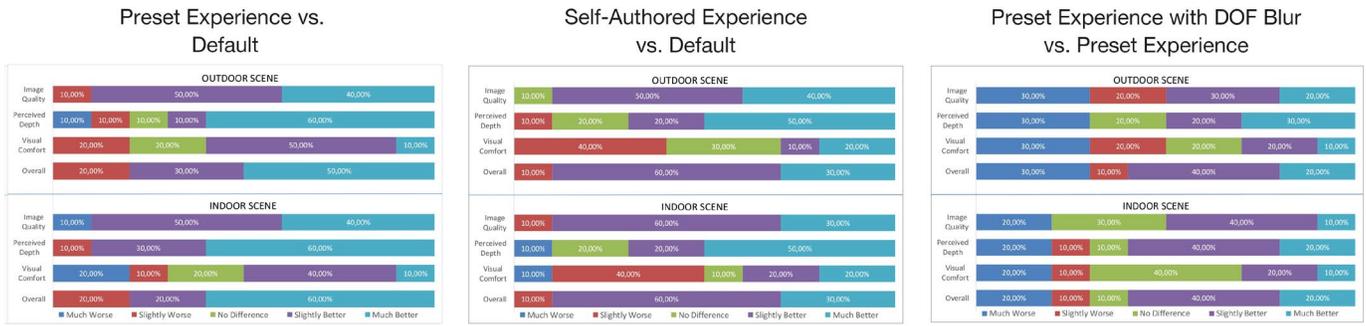

Fig. 11: Relative user preferences collected with the questionnaires are given in percentages. Scores are relative to the compared setting.

### 5.7 Results

Figure 10 illustrates the results of the individual session evaluations for image quality, depth perception and visual comfort for all three configurations. The user preferences in percentages in the three configurations are given in Figure 11.

For the first two configurations where our approach is evaluated against the Default Experience setting, it is seen that, in both Preset Experience and Self-Authored Experience settings, our approach was rated significantly higher on average in terms of image quality and perceived depth in both scenes. 9 subjects, for both settings and both scenes, indicated their preference of our approach to the Default Experience in terms of image quality. Similarly, in the outdoor scene, 7 subjects preferred our approach to the Default Experience in terms of perceived depth with both the Preset and the Self-Authored Experiences. For the indoor scene, 7 subjects favored the Preset Experience and 9 subjects favored the Self-Authored Experience in terms of perceived depth.

In terms of visual comfort, again for the first two configurations, our approach was rated only slightly better in the outdoor scene with both settings. In the indoor scene, however, it was rated slightly worse with the Preset Experience and the same with the Self-Authored Experience. It is seen that while the Default Experience was favored in terms of visual comfort to the Preset Experience by 2 and 3 of the subjects in the outdoor and indoor scenes, respectively, 4 and 5 of the subjects preferred it to their own Self-Authored Experiences in the outdoor and indoor scenes, respectively. The subject with VR application development background, on the other hand, was among the group who found their own Self-Authored Experience more visually comforting. These findings imply that novice VR users may still have a hard time when they first start authoring VR experiences for their ideal visual comfort level, however easy the interface is to grasp and use.

The questionnaire item that queries the overall preference of the subjects garnered responses demonstrating that our approach was considerably well-received with both the Preset Experience and the Self-Authored Experience settings. The breakdown of the results show that while the Self-Authored Experience was preferred by more subjects (9 in both scenes) than the Preset Experience (8 in both scenes), more subjects regarded the Preset Experience as "much better" with respect to the Default (5 in the outdoor scene and 6 in the indoor scene) than they did the Self-Authored Experience (3 in both scenes).

The results for the third configuration, which facilitates to evaluate the impact of our DoF blur implementation in combination with our approach, show that the combination was generally found to degrade both image quality and visual comfort. While, more subjects indicated their overall preference towards the session with DoF blur, the average ratings of the session without DoF blur are higher in all three evaluation criteria for the outdoor scene and similar for the indoor scene.

Figure 12 presents a sample of *depth charts* that contrast the depth distributions resulting from our approach with the results of the default arrangement in the two scenes. That is, the pair of charts for each scene is obtained by following the same camera path using the default settings once and once using our approach. On the charts, minimum and maximum depth values along the camera path are given with respect to the HMD display. It is seen that while the default arrangement constrains the zero-parallax plane to a fixed short distance to the stereo camera pair leading to a very narrow band of negative disparity region, our approach allows a dynamic yet smooth depth narrative in both negative and positive disparity regions.

### 6 CONCLUSION

In this paper we have described an effective method of real-time automated stereoscopic camera control in VEs towards providing users with impressive VR experiences that are rich in depth. We have also introduced an immersive design interface for authoring unique VR experiences to be used with the proposed method. We believe both of these novel contributions that are ready to be used with existing consumer HMDs will stimulate new directions in stereoscopic rendering research. Moreover, we have addressed the issues with the stereoscopic parameter extrema that may cause undesired effects and proposed a dynamic limiting scheme.

The results of the user evaluation study demonstrate that our method is able to enhance user experience, as intended, especially in terms of overall perceived depth and image quality. The method is also seen to slightly improve the visual comfort in the scene with wider depth range and to keep it at similar levels in the scene with narrower depth range. The use of DoF blur added to our proposed method did not help to improve the experience further, since the majority of the user assessments were in accord with some earlier studies [5, 9, 10]. Surely, a further standalone study with higher number of configurations and larger sample size in which users can be grouped as first time VR users, VR enthusiasts and VR developers would be beneficial for a more thorough analysis.

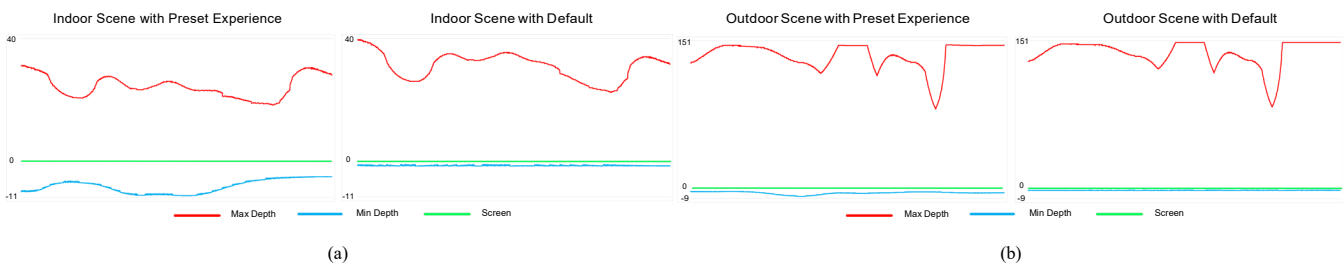

Fig. 12: Relative maximum and minimum depth value chart of (a) the indoor scene, (b) the outdoor scene.

While it is observed that using our approach in an interactive setting where the user can roam the scene freely without being restricted to the camera path set in the selected depth layout lead to acceptable results, we believe further improvements to the method are crucial for proper adaptation to use in interactive VR applications.